\documentclass[aps,prl,showpacs,twocolumn,amsmath,amssymb]{revtex4-1}
\usepackage{graphicx}% Include figure files
\usepackage{units}
\usepackage{upgreek}
\usepackage{color}
\usepackage{hyperref}
\usepackage[all]{hypcap}
\usepackage{braket}
\usepackage[left]{lineno}

\preprint{APS/123-QED}

\newcommand{\cev}[1]{\reflectbox{\ensuremath{\vec{\reflectbox{\ensuremath{#1}}}}}}

\begin{document}
% Show text/columnwidth, usepackage layouts
%\printinunitsof{inch}
%Columnwidth \prntlen{\columnwidth}\\
%Textwidth \prntlen{\textwidth}\\
%Textheight \prntlen{\textheight}\\

\title{Rydberg molecule-induced remote spin-flips}
\author{Thomas Niederpr\"um$^1$}
\author{Oliver Thomas$^{1,2}$}
\author{Tanita Eichert$^1$}
\author{Herwig Ott$^1$}
\email{ott@physik.uni-kl.de}
\affiliation{$^1$ Research Center OPTIMAS, Technische Universit\"at Kaiserslautern, 67663 Kaiserslautern, Germany}
\affiliation{$^2$Graduate School Materials Science in Mainz, Staudinger Weg 9, 55128 Mainz, Germany}

\begin{abstract}

We have performed high resolution photoassociation spectroscopy of rubidium ultra long-range Rydberg molecules in the vicinity of the 25$P$ state. Due to the hyperfine interaction in the ground state perturber atom, the emerging mixed singlet-triplet potentials contain contributions from both hyperfine states. We show that this can be used to induce remote spin-flips in the perturber atom upon excitation of a Rydberg molecule. When furthermore the spin-orbit splitting of the Rydberg state is comparable to the hyperfine splitting in the ground state, the orbital angular momentum of the Rydberg electron is entangled with the nuclear spin of the perturber atom. Our results open new possibilities for the implementation of spin-dependent interactions for ultracold atoms in bulk systems and in optical lattices.
\end{abstract}

\pacs{32.80.Ee, 31.15.ae, 34.50.Cx}
\maketitle

%===============================================================================
% Introduction %
%===============================================================================

Implementing tunable short-range interactions in ultracold quantum gases has proven to be key to study quantum phase transitions \cite{Greiner2002} and strongly interacting many-body systems \cite{Paredes2004,Zwierlein2012}. 
The most commonly used techniques are magnetic Feshbach resonances \cite{Chin2010} and confinement-induced effective interactions \cite{Olshanii1998,Dalibard2011}. 
Long-range interactions beyond the pure contact interaction are more challenging to achieve. 
Possible realizations include second order tunneling \cite{Foelling2007}, cavity-mediated interactions \cite{Baumann2010}, magnetic dipolar interactions in high spin atomic species \cite{Griesmayer2005,Lev2011,Ferlaino2012} and electric dipolar interactions between heteronuclear molecules \cite{Ni2009}. 
Exciting atoms to Rydberg states is another way to induce long-range interactions, as evidenced by the demonstration of the Rydberg blockade \cite{Lukin2001,Singer2004,Tong2004,Gaetan2009} and anti-blockade \cite{Amthor2010,Weber2015}. 
Recently, these concepts were transferred to the realm of ultracold quantum gases \cite{Schauss_2012}. 
First experimental results with off-resonant excitation schemes show that for short times, coherent interactions between ground state atoms can be generated \cite{Zeiher2016}. 
In most such ''Rydberg dressing'' schemes the interaction is based on admixing Rydberg excitations to two particles, resulting in energy shifts which scale quadratically with the driving laser intensity. 
This narrows the parameter window for coherent effects drastically \cite{Zeiher2016,Rolston2010}.

The discovery of Rydberg macrodimers \cite{Boisseau2002,Overstreet2009} and Rydberg molecules \cite{Bendkowsky2009} has opened up an increasing field of research, combining ultracold chemistry with many-body physics and low energy electron scattering. 
Rydberg molecules are bound by the contact interaction between the Rydberg electron and a ground state perturber atom. 
The large extension of the Rydberg electron wave function ($50 - 1000\,$nm) makes it possible to induce long-range interactions between two spatially separated (remote) ground state atoms that otherwise interact solely through contact interaction on a typical length scale of $5\,\mathrm{nm}$ in the case of rubidium.
In contrast to the usual Rydberg dressing of single species gases\cite{Rolston2010}, only one excitation is required, thus leading to a more favorable first order process, which scales linearly with the laser intensity. 

For alkali atoms, one can distinguish three different types of molecules: ultra-long range Rydberg molecules \cite{Greene2000,Bendkowsky2009}, trilobite molecules \cite{Booth2015} and butterfly molecules \cite{Chibisov2002,Hamilton2002,Niederpruem2016}. 
While sharing a similar binding mechanism, they differ in the degree of perturbation, which is imposed by the ground state perturber to the Rydberg electron wave function.
Here, we change the perspective and study the effect of the binding mechanism on the perturber atom.
In agreement with theoretical predictions we experimentally confirm the presence of spin-flip processes in the ground state perturber upon excitation of ultra long-range Rydberg molecules. We can also excite particular Rydberg states, where the Rydberg orbital angular momentum is strongly entangled with the nuclear spin of the perturber atom. 
For the 25$P$ state of rubidium, both effects are active over a distance of up to 50\,nm between the two atoms.
As we use a single photon excitation scheme to excite the molecules, we avoid spontaneous scattering from an intermediate level. Our technique is therefore suited to induce coherent and dissipative interactions in ultracold atomic gases.
This includes the realization of optical Feshbach resonances \cite{Nicholson2015} involving Rydberg molecules and spin-dependent dissipative processes.

\begin{figure}[t]
\begin{center}
\includegraphics[width=\columnwidth]{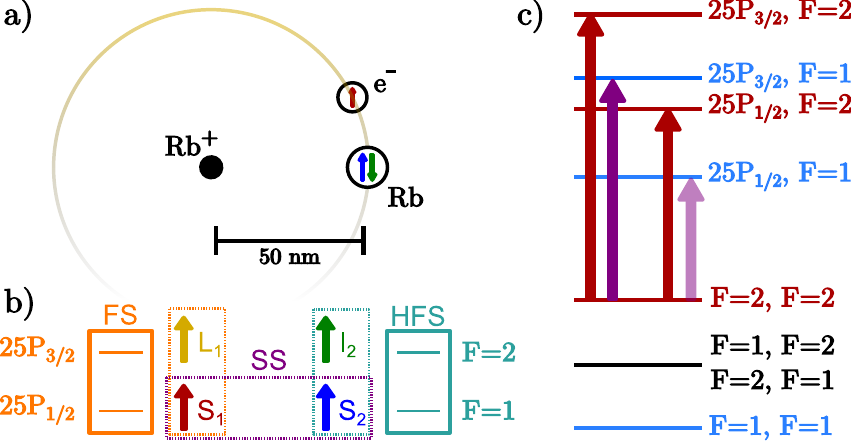}
\end{center}
\caption{(Color online). (a) The contact interaction between the Rydberg electron $e^-$ and the ground state atom $\mathrm{Rb}$ leads to a spin-dependent interaction over distances up to \unit[50]{nm} in the $25P$ state. (b) The angular momentum coupling scheme shows how the spin-spin interaction (SS) couples the fine structure (FS) of the Rydberg atom with the hyperfine structure (HFS) of the perturber. The color of the arrows corresponds to the colors used in a).  (c) Transition scheme. When the sample is in the $F=2$ ground state only the atomic transitions to states adiabatically connecting to $F=2$ states are possible (red arrows). Due to the hyperfine mixing of the molecular interaction (see text), also transitions to molecular states in the $F=1$ spectrum are possible (purple arrows)} 
\label{fig:excitationScheme}
\end{figure}

\begin{figure*}[t]
\begin{center}
\includegraphics[width=\textwidth]{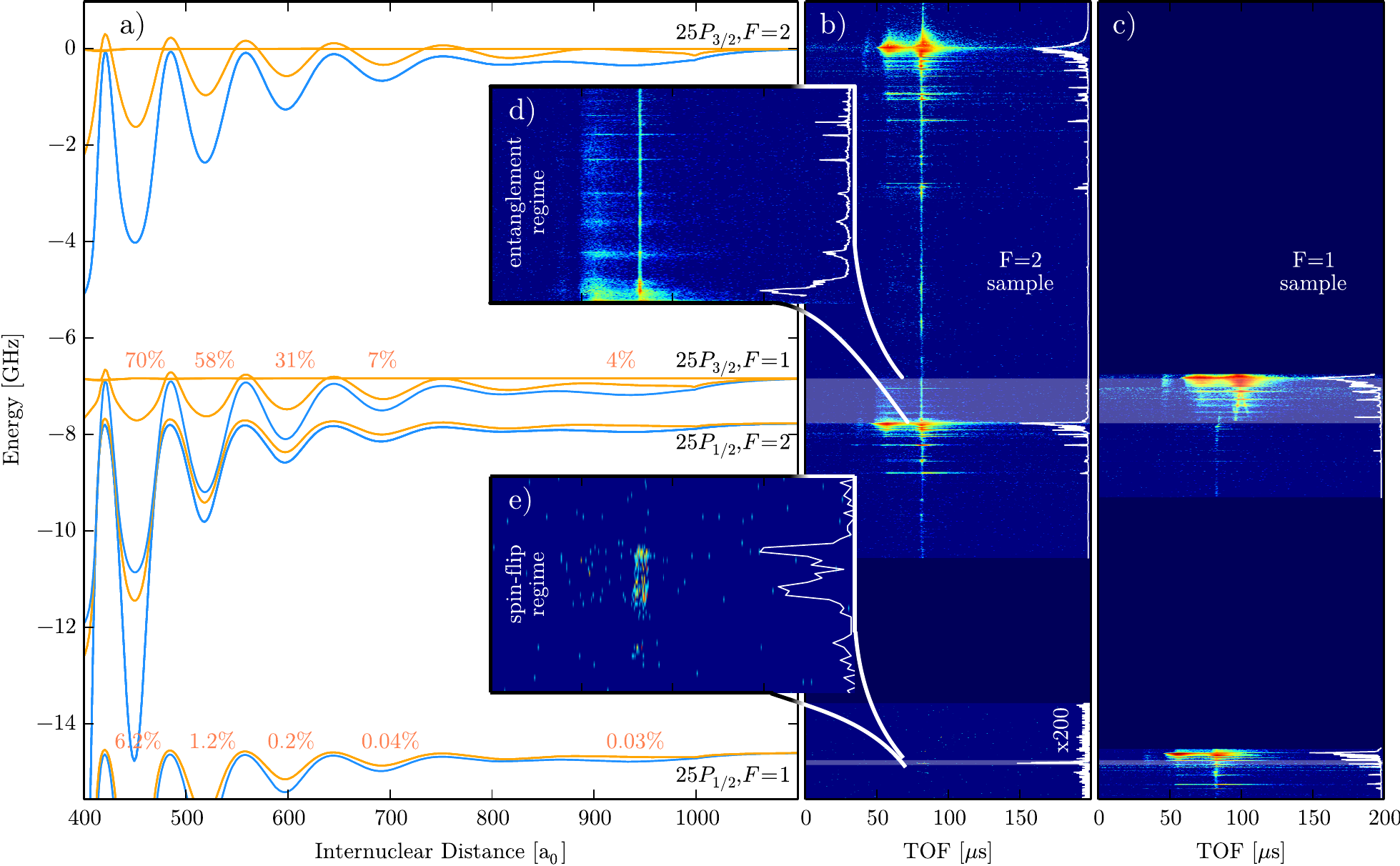}
\end{center}
\caption{(Color online). (a) Adiabatic potential energy curves (PECs) for ultra-long range Rydberg molecules of rubidium 87. The PECs adiabatically connect to the four different $25P$ states. The blue PECs are of pure triplet type and do not mix the hyperfine states, the orange PECs are of mixed singlet-triplet character and contain both hyperfine states. The red numbers give the admixture of opposite spin to the states in the respective wells of the potential. (b) and (c) show the measured TOF spectrum for a BEC prepared in a pure $F=2$ and $F=1$ state, respectively.  The time of flight is \unit[58]{$\mu s$} for the $Rb^+$ ions and \unit[82]{$\mu s$} for the $Rb_2^+$. The deep blue regions were not measured. The white lines show the flight-time integrated spectrum. The insets (d) and (e) show zoom-ins on highlighted parts of the $F=2$ spectrum, illustrating the entanglement and the spin-flip regime, respectively.} 
\label{fig:TheoCompareSpectrum}
\end{figure*}

%===============================================================================
%Model
%===============================================================================
Based on Fermi's original idea of $s$-wave scattering by a quasi free electron \cite{Fermi1934}, the interaction of a ground state perturber atom at a distance $R$ inside the Rydberg wave function is described by a zero range pseudo-potential $V_s(R) = 2\pi A_s(k_R)\delta(r-R)$ with a scattering length $A_s$ that depends on the classical electron momentum $k_R = \sqrt{\frac{2}{r}-\frac{1}{n_\mathrm{eff}^2}}$. Due to shape resonances appearing in the low energy scattering of electrons and alkali atoms, it is crucial to extend the pseudo-potential to also include the $p$-wave scattering process \cite{Omont1977}. Taking further into account the different scattering length for singlet and triplet scattering and the hyperfine structure in the perturber atom, the Hamiltonian for the molecular system in Born-Oppenheimer approximation reads \cite{Anderson2014}

\begin{align}
\hat{H} &= \hat{H}_0 \nonumber\\
&+ 2\pi \left[a_s^S(k_R)\hat{\mathbb{I}}^S + a_s^T(k_R)\hat{\mathbb{I}}^T\right]\delta^{(3)}(\vec{r}-\vec{R}) \nonumber \\
&+ 6\pi \left[a_p^S(k_R)\hat{\mathbb{I}}^S + a_p^T(k_R)\hat{\mathbb{I}}^T\right]\delta^{(3)}(\vec{r}-\vec{R})\frac{\cev{\nabla}\cdot\vec{\nabla}}{k_R^2} \nonumber \\
&+ A\hat{\vec{S}}_2\cdot\hat{\vec{I}}_2.
\label{eq:Hamiltonian}
\end{align}

\noindent
Here, $\hat{H}_0$ is the atomic Hamiltonian of the Rydberg atom including the fine structure, $a_s^S$ ($a_s^T$) are the $s$-wave scattering lengths for singlet (triplet) scattering and  $a_p^S$ ($a_p^T$) the $p$-wave scattering lengths for singlet (triplet) scattering. 
The projector on the triplet subspace can be expressed in terms of the spin $\hat{\vec{S}}_1$ of the Rydberg electron and the spin $\hat{\vec{S}}_2$ of the perturber and is given by $\hat{\mathbb{I}}^T = \hat{\vec{S}}_1\cdot\hat{\vec{S}}_2 + 3/4$.
The singlet projector is $\hat{\mathbb{I}}^S = 1 -\hat{\mathbb{I}}^T $.
The hyperfine coupling in the perturber is described by the hyperfine constant $A = 3.4\,\mathrm{GHz}$ (for $^{87}\mathrm{Rb}$) and the coupling between the electronic spin $\hat{\vec{S}}_2$ and the nuclear spin $\hat{\vec{I}}_2$ of the perturber.
Fig.\,\ref{fig:excitationScheme} shows the different angular momentum couplings that occur in the Rydberg molecules.
%Note that the full interaction Hamiltonian (1) is not diagonal in any of the usual atomic angular momentum eigenstate basis. 

In order to calculate the Born-Oppenheimer potential energy curves (PEC), we have carried out a full diagonalization of the Hamiltonian \eqref{eq:Hamiltonian}
%with a truncated basis set of the form $|n,l,j,m_j\rangle \otimes |m_s, m_I\rangle$, using the usual $n$, $l$, $j$, $m_j$ quantum numbers of the spin-orbit coupled Rydberg state and the $m_s$, $m_I$ quantum number of the perturber electron and nuclear spin, respectively.
%For the following discussion however, we will use the notation based on the hyperfine state of the perturber atom, e.g., $25P_{1/2};F=2$.
and the resulting eigenenergies as a function of the internuclear distance $R$ are shown in Fig.~\ref{fig:TheoCompareSpectrum}a.
The corresponding excitation scheme including the initial state of the two atoms is shown in Fig.\,\ref{fig:excitationScheme}c.
As a consequence of the hyperfine interaction in the perturber atom, the singlet- and triplet states are mixed and the Hilbert space can no longer be separated into the according subspaces.
The emerging eigenenergies therefore feature one pure triplet potential energy curve (blue lines in Fig.~\ref{fig:TheoCompareSpectrum}a) and one of mixed singlet-triplet  character (orange lines) \cite{Anderson2014, Sasmannshausen2015, Boettcher2016}.
This argument also applies the other way around: due to the singlet- and triplet terms in the Hamiltonian, the subspaces of the $F=1$ and the $F=2$ hyperfine states of the perturber are mixed and thus the mixed character PEC contains both hyperfine states.
The degree of hyperfine mixing depends on the relative strength of the Rydberg - ground state interaction with respect to the hyperfine interaction and accordingly we can identify two different regimes.
In the more general case, which we denote as spin-flip regime, the interaction is small compared to the hyperfine splitting and thus the admixture of the opposite hyperfine state is small.
In the system at hand, this situation is realized for the $25P_{3/2}; F=2$ and the $25P_{1/2}; F=1$ states. For the latter, the molecular states have the form:
\begin{equation}
\ket{\Phi}_\mathrm{sf} = \alpha\ket{25P_{1/2}}\ket{F=1} + \epsilon \ket{25P_{1/2}}\ket{F=2}+...
\end{equation}
with $\alpha \approx 1$ and $\epsilon\ll 1$. The contribution of other states is of similar magnitude.
Starting from two atoms in the $F=2$ state, the small admixtures $\epsilon$ allow for the coupling to Rydberg molecules of opposite ground state spin and the excitation process can be seen as a spin-flip collision between the Rydberg electron and the ground state perturber.
Since the created molecule is predominantly in the flipped spin state, the perturber atom will most likely pertain its flipped spin state, even upon spontaneous decay of the molecular state. 

A peculiar second regime appears for the asymptotic free $25P_{1/2} ; F=2$ and  $25P_{3/2}; F=1$ states.
Since the spin-orbit splitting of the $25P$ state almost equals the hyperfine splitting of the perturber, these two levels are separated by only $929\,\mathrm{MHz}$, which is comparable to the Rydberg-ground state interaction energy at small internuclear distances.
Consequently the theory predicts strong mixing up to 50\% of the two states.
For large internuclear distances, we still find an admixture of a few percent, even in the outermost well (Fig.~\ref{fig:TheoCompareSpectrum}a).
%A much stronger mixing in excess of 50\% appears at the $25P$ state since it's spin-orbit splitting is almost equal to the hyperfine splitting of the perturber and thus the asymptotic free $25P_{1/2} ; F=2$ state and the $25P_{3/2}; F=1$ state are separated by only $929\,\mathrm{MHz}$.
These states are predominantly a superposition of the two asymptotic ones,
\begin{equation}
\ket{\Phi}_\mathrm{ent} = a \ket{25P_{1/2}}\ket{F=2} + b \ket{25P_{3/2}}\ket{F=1},
\end{equation}
with $a,b\approx 0.1...0.8$, which entangle the fine structure state of the Rydberg atom with the hyperfine state of the perturber. They are distinct from the spin-flip regime by the much stronger mixing. We denote this regime as the ''entanglement'' regime.

%This coupling ultimately allows to change the hyperfine state of the perturber atom upon transition into the molecular state.
%This becomes especially evident in the case of the $25P$ state in $^{87}Rb$ since the hyperfine splitting of the perturber and the spin-orbit splitting of the Rydberg state are almost equal.
%Consequently, we see a strong mixing between the asymptotic free $25P_{1/2} \otimes 5S_{1/2}, F=2$ state and the $25P_{3/2} \otimes 5S_{1/2}, F=1$ state, as indicated in Fig.~\ref{fig:TheoCompareSpectrum}a. 

%===============================================================================
% Experimental Setup and Sequence %
%===============================================================================
In order to experimentally prove the existence of hyperfine mixing in Rydberg molecules, we photoassociate ultra-long range Rydberg molecules in the vicinity of the $25P$ state from a Bose-Einstein condensate (BEC) of $^{87}\mathrm{Rb}$.
The experimental apparatus is described in detail in reference \cite{Manthey2014}.
In brief, a BEC of $10^5$ atoms and a temperature of $100\,\mathrm{nK}$ is prepared in a crossed YAG optical dipole trap by forced evaporation to final trapping frequencies of $2\pi \times 67\,\mathrm{Hz}$ in all three directions.
Due to a small magnetic field gradient present during evaporation, the BEC is spin polarized in the $5S_{1/2}, F=1, m_F=+1$ ground state.
Using microwave radiation the spin state of the atoms can be transferred to the fully stretched $F=2, m_F=+2$ state with a Landau-Zener sweep at a fidelity of close to 100\%.
The photoassociation of Rydberg molecules is achieved by a frequency doubled cw dye laser at a wavelength of $297\,\mathrm{nm}$ and a laser linewidth below $700\mathrm{kHz}$. 
Once produced, the Rydberg molecules can decay into ions either by photoionization, leading to a $\mathrm{Rb}^+$ atomic ion, or by associative ionization, leading to a $\mathrm{Rb}_2^+$ molecular ion\cite{Niederpruem2015,Schlagmueller2016}.
The experimental sequence consists of 1000 excitation pulses ($1\,\mathrm{\mu s}$) with subsequent continuous ion detection ($200\,\mathrm{\mu s}$).
Due to the different mass the atomic and molecular ions have different time of flights (TOF) to the ion detector. From the decay of the signal, we can additionally extract the lifetime of the produced molecules.
We have performed photoassociation spectroscopy with a resolution of 1\,MHz, spanning more than 10\,GHz. 

\begin{figure}[t]
\begin{center}
\includegraphics[width=\columnwidth]{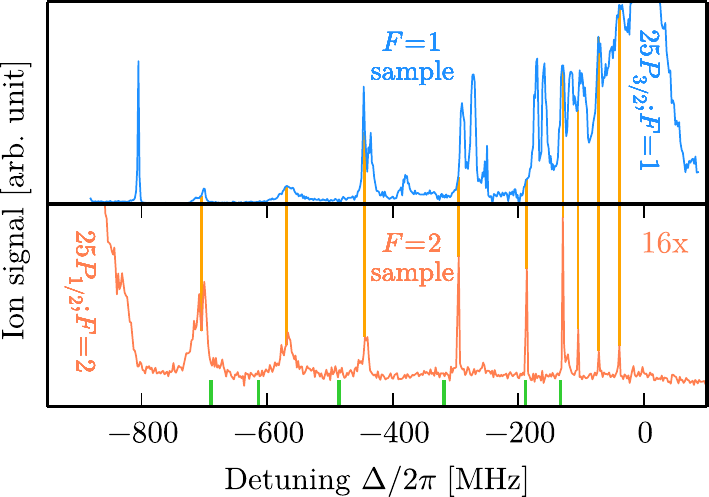}
\end{center}
\caption{(Color online). Comparison of the molecular spectra in the region between the $25P_{3/2}; F=1$ state (\unit[0]{MHz}) and the $25P_{1/2}; F=2$ state ($\unit[-929]{MHz}$) in a sample with all atoms in the $F=1$ state (blue) and all atoms in the $F=2$ state (red). Since in the $F=2$ spectrum we only observe the mixed type PEC (see text) that also appears in the $F=1$ spectrum, every line in the $F=2$ spectrum has a corresponding line in the $F=1$ spectrum (orange lines). The calculated energies of the lowest bound states in each well of the mixed type PEC (green bars) agree with the observed resonances. Compared to the $F=1$ spectrum the $F=2$ spectrum is magnified by a factor of 16. Due to an uncertainty in the frequency calibration, the $F=2$ measurement was stretched by 2\%, in accordance with the frequency mismatch observed in comparable measurements.} 
\label{fig:F1F2Compare}
\end{figure}

%In order to have frequency markers in our spectrum, we employ residual light from the 0th order of the switching AOM, that constantly illuminates the sample with a frequency offset of \unit[211]{MHz}. 
%For strong spectroscopic lines we thus see a weak ghost signal in the TOF spectrum that is constant in time and shifted in frequency. Wherever the molecular lines are strong enough we use these markers to correct the relative frequency from the atomic resonance.
%\textcolor{red}{We estimate an overall absolute accuracy of the experimental spectrum of better than xx\,MHz.}  
The full spectrum for a BEC in the $F=2$ state is shown in Fig.~\ref{fig:TheoCompareSpectrum}b along with the relevant parts of the spectrum for a BEC in the $F=1$ state in Fig.~\ref{fig:TheoCompareSpectrum}c.
The most prominent features in each spectrum are the two bare atomic transitions, which can only appear for those states that match the hyperfine state of the prepared BEC.
As discussed above this does no longer hold for the molecular states.
Instead, some of the molecular lines of the $25P_{3/2}; F=1$ spectrum appear also on the blue side of the $25P_{1/2}; F=2$ state in the $F=2$ spectrum  (Fig.~\ref{fig:TheoCompareSpectrum}d).
As those lines can only originate from the mixed type potential, we compare in Fig.~\ref{fig:F1F2Compare} the lines in the $F=2$ spectrum to the calculated energies of the lowest bound states in each well of the mixed type potential adiabatically connecting to the $25P_{3/2}; F=1$ state (green bars).
The three highest energy lines can not be attributed to a ground state in any of the wells and are probably higher excited states.
The residual six observed lines coincide with the predicted bound state energies within 10\%.
Even stronger evidence for those resonances originating from the hyperfine mixing in the $25P_{3/2}; F=1$ potential curve arises from the direct comparison of both spectra (Fig.~\ref{fig:F1F2Compare}).
Except for the line close to $-200\,\mathrm{MHz}$ it is possible to attribute each line in the $F=2$ spectrum to a corresponding line in the $F=1$ spectrum.
This not only provides strong evidence for the discussed hyperfine mixing but also allows us to identify which peaks in the $F=1$ spectrum belong to the mixed potential and which, by exclusion, belong to the triplet potential.
%The lines in the $F=2$ spectrum furthermore agree within 10\% with the calculated energies of the lowest bound states in each well of the mixed type potential adiabatically connecting to the $25P_{3/2}; F=1$ state.
%This not only proofs that these molecular states are a mixture of hyperfine states but also allows us to identify which peaks in the $F=1$ spectrum belong to the mixed potential and which, by exclusion, belong to the triplet potential.
The comparable magnitude of the three interactions which couple the different angular momenta (Fig.\ref{fig:excitationScheme}b) leads to the before mentioned entanglement between the orbital degree of freedom of the Rydberg electron and the nuclear spin of the ground state atom.
Since the interaction between two such molecular entangled states depends on the fine structure state of the Rydberg atom this can be used to entangle the spin of ground state atoms over the typical length scale of Rydberg-Rydberg interactions.

In contrast to the strong lines in the entanglement regime it is more challenging to observe the hyperfine mixing in the spin-flip regime.
Due to the high signal-to-noise ratio provided by the ion signal we are nevertheless able to see a molecular line at $-14.8\,\mathrm{GHz}$ in the $F=2$ spectrum (Fig.~\ref{fig:TheoCompareSpectrum}e), which, in comparison with the calculated PECs, can only be attributed to the $25P_{1/2}; F=1$ state (highlighted region in Fig.~\ref{fig:TheoCompareSpectrum}b).
Since the observed line differs only by \unit[32]{MHz} from the expected energy of the lowest bound state in the well at \unit[692]{$\mathrm{a_0}$} of the $25P_{1/2}; F=1$ mixed potential, we assume the admixture of the $F=2$ state to be on the order of $\epsilon^2 =0.04\,\%$.
The detection of bound states at higher internuclear distances is hindered by the small hyperfine mixing. At closer distances on the other hand the reduced probability to find a pair of atoms decreases and pushes the line strength below our detection limit.
It should be noted that the observed molecular line has the lowest energy of all possible transitions depicted in Fig.~\ref{fig:excitationScheme} and thus the presence of $F=1$ atoms in the initial sample can not explain the observed signal.
We have thus experimentally shown a spin-flip of the ground state perturber upon photoassociation of a Rydberg molecule over a distance of 35\,nm between the two atoms.

Due the high particle density and the presence of collective modes in the BEC, many-body effects beyond the two-particle picture might influence the observed spin-flip mechanism. 
However, the possibility to spectroscopically address a well-defined molecular state allows us to selectively photoassociate only atom pairs that don't have any additional ground state atom inside the Rydberg wave function. 
Furthermore, bound states of two or more perturber atoms \cite{Gaj2014} are strongly suppressed due to the geometric constraints imposed by the p-state wave function \cite{Sasmannshausen2015}. 
Also, the molecular formation process can hardly excite collective modes in the BEC \cite{Karpiuk2015,Wang2015} as the size of the molecules is much smaller than the healing length $\xi = 230\,\mathrm{nm}$.”

%===============================================================================
%Discussion
%===============================================================================

%We close this letter discussing several applications of the observed Rydberg molecules in ultracold quantum gases.
When the separation between the two atoms in the molecular state is much smaller than the typical interparticle distance in a quantum gas or in an optical lattice, the resulting interaction might still be classified as ''short-range''.
It can then be used to modify the contact interaction between the atoms. In fact, optical Feshbach resonances are based on the coupling of a free two-particle scattering state to a molecular bound state with a photoassociation laser.
Due to intrinsic losses, molecular states with long lifetimes and minimal off-resonant scattering from the bare atomic resonance are mandatory to apply this concept.
Experiments have so far been performed on different atomic species, most promising results have been obtained for ytterbium and strontium \cite{Nicholson2015}.
The latter features a molecular decay rate of $\gamma/2\pi=14\,$kHz \cite{Nicholson2015}. While a shift in the scattering length could be demonstrated successfully, losses still pose a serious challenge.
With the presented spin-flip mechanism, Rydberg molecules can overcome these limitations, due to the absence of scattering from a bare atomic resonance.
Since furthermore the decay rates of $\gamma/2\pi=10 - 30$\,kHz (extracted from the time of flight spectra in Fig.\,\ref{fig:TheoCompareSpectrum}b) are compatible, we speculate that the coupling to spin-flipped Rydberg molecules is in principle suited to implement optical Feshbach resonances without scattering from a nearby bare atomic resonance.
%With sufficient coupling strength the presented spin-flip mechanism can overcome these limitations when it is employed to off-resonantly excite Rydberg p-state molecules with comparable decay rates of $\gamma/2\pi=10 - 30$\,kHz (extracted from Fig.\,\ref{fig:TheoCompareSpectrum}b) and is therefore a possible way to implement optical Feshbach resonances without scattering from a nearby bare atomic resonance (Fig.\,\ref{fig:excitationScheme}).
%For Rydberg molecules, we extract from the time of flight spectra (Fig.\,\ref{fig:TheoCompareSpectrum}b) similar molecular decay rates, $\gamma/2\pi=10 - 30$\,kHz, being in the order of the natural lifetime of the Rydberg state.

%Looking at the level scheme in Fig.\,\ref{fig:excitationScheme}, it becomes clear that such Feshbach resonances could be implemented in a spin-resolved fashion.
%Provided enough coupling strength can be achieved, the use of spin-flip molecules is especially interesting, as it completely avoids the scattering from the bare atomic resonance.

For the resonant excitation of Rydberg molecules, non-unitary time evolution occurs. Upon excitation, spontaneous decay of the Rydberg molecules and associative ionization \cite{Niederpruem2015} lead to the loss of one or both atoms. However, these losses occur only for the addressed combination of hyperfine states (Fig.\,\ref{fig:excitationScheme}c). In an optical lattice with a two component quantum gas, one could therefore induce losses in doubly occupied sites with a specific spin composition. The phase space dynamics can then drive the system in a correlated spin state, which is decoupled from the loss process.

%The potential of inducing coherent long-range interactions in optical lattices cannot straightforward be realized with ordinary ultra-long range molecules. As the binding energy of the Rydberg molecules scales as $n^{-6}$, where $n$ is the main principal quantum number, the expected binding energy at a molecular bond length of 260\,nm, (corresponding to a main principal quantum number of $n=70$) is only 100\,kHz. However, chosing a trilobite state by exciting Rydberg f-states with a three-photon transition might overcome this problem, thus enabling nearest-neighbor interactions in optical lattices. 

%===============================================================================
%Conclusion
%===============================================================================

In conclusion, we have performed high resolution photoassociation spectroscopy of p-state Rydberg molecules and have demonstrated spin-flip collisions in Rydberg molecules. In our case these spin-flip processes happen for an interatomic distance of about 35\,nm. We also resolve molecular states, which feature strong entanglement between the orbital angular momentum of the Rydberg electron and the nuclear spin of the ground state perturber atom. Our results point at the possible realization of optical Feshbbach resonances employing Rydberg molecules and provide new means to induce unitary and non-unitary interactions in ultracold quantum gases. This approach works for all atomic species or mixtures which support Rydberg molecules.

% Acknowledgements %
We thank C. Greene and J. P\'erez-R\'ios for valuable discussions about the spin flip processes and the potential energy curves. We thank C. Lippe for discussing the manuscript and I. Fabrikant for providing the Rb-$e^-$ scattering phase shifts.
We acknowledge financial support by the DFG within the SFB/TR49. O.T. is funded by the graduate school of excellence MAINZ.

\bibliographystyle{apsrev4-1}

%\begin{thebibliography}{999}

%\bibitem{Gosh2010}
%S. Ghosh, {\it et al.} Nature Mat. {\bf 9}, 555 (2010).

%\end{thebibliography}

\bibliography{spinflip_references}

\end{document}